# Heavy Ion Beam in Resolution of the Critical Point Problem for Uranium and Uranium Dioxide

I. Iosilevski, V. Gryaznov

*Moscow Institute of Physics and Technology, State University (Russia)*
*Institute of Problems of Chemical Physics RAS (Russia)*

Heavy Ion Beam (HIB) irradiation of matter have an important advantage in comparison with other (traditional) sources of energy deposition – laser heating, electron beam, electrical discharge etc. – high penetration length (~ 10 mm) in cold condensed matter. This property of HIB gives an extraordinary chance to reach the *uniform heating* regime when HIB irradiation is being used for thermophysical property measurements. Perspective of such HIB application moves one to revise the priorities in HIB development: preferable energy levels, beam-time duration, beam focusing, deposition of the sample etc. [1].

## The Problem of Critical Point Parameters Estimation

The high energy density equation of state (EOS) of metals, including parameters of their gas-liquid coexistence up to the critical point (CP), is first needed as an important ingredient in development of perspective energetic devices, including the inertial fusion (IF). One's interest in properties of metals near their CP is quickened also by presumption of possible thermodynamic anomalies like hypothetical Plasma Phase Transition (PPT) [2, 3] or by theoretical predictions of violation of positiveness for $(\partial P/\partial T)_V$ [4] *etc*. The high-temperature EOS of metallic uranium as well as that for uranium-bearing compounds like $UO_{2\pm x}$, $UF_6$ *etc.*, are also of principal importance in analysis of hypothetical *severe reactor accidents* at presently exploited nuclear plants (see for example [5, 6]) as well as in development of perspective non-traditional schemes of fission energy devices (for example, *gas-core nuclear reactor* [7] *etc*).

Critical temperature and pressure for most of metals are too high for precise experimental study (except for heavy alkali and mercury). Consequently the CP parameters (CPP) for majority of metals, as well as CPP of $UO_2$, are presently known mostly from theoretical predictions. The dominating approach in such estimations is based on presumption of strong inherent correlation between CPP and low-temperature properties of condensed phase. So far in practice it takes form of *far extrapolation* of these known low-temperature properties. An alternative approach makes accent on "plasma" features of behavior for majority of metals in vicinity of their critical point. As result, the recommended procedure of CPP estimation [8, 9] based on correlation of metal CPP with ionization potential and effective valence.

As a rule all the approaches give rather close results for CPP-s of many substances. Uranium is an extraordinary example (not the only one) of *strong contradiction* between different ways of such estimations. Thus the dominating approach is based on fundamental *caloric quantity* – cohesive energy. It leads to relatively *high values* of predicted critical temperature and pressure: $T_c^{(U)} \sim 12 \div 13 \cdot 10^3$ K; $P_c^{(U)} \sim 600 \div 900$ MPa [10–14]. Another widely accepted approach based on *thermal quantity* – dependence of experimentally measured liquid density on temperature – $\rho_{liquid}(T)$. It leads to surprisingly *low values* of predicted uranium critical temperature $T_c$ ($T_c^{(U)} \sim 6 \div 7 \cdot 10^3$ K) [15–17]). At last, the "plasma approach" [8, 9] gives *intermediate* level of predicted $T_c$ ($T_c^{(U)} \approx 9 \cdot 10^3$ K). The discussed situation is illustrated at Fig.1.

The discrepancy between CPP-s estimated from the low-temperature thermal and caloric properties [10–17], or from "plasma" arguments [8, 9], becomes absolutely *incompatible* when being considered jointly with the *third* source of low-temperature empiric information on thermodynamic property of condensed phase – Gibbs free energy of liquid. The latter is incorporated into the CPP problem through the far extrapolation of low-temperature saturated pressure dependence, $P_s(T_s)$. The present uncertainty of such extrapolation for uranium is illustrated at Fig. 2

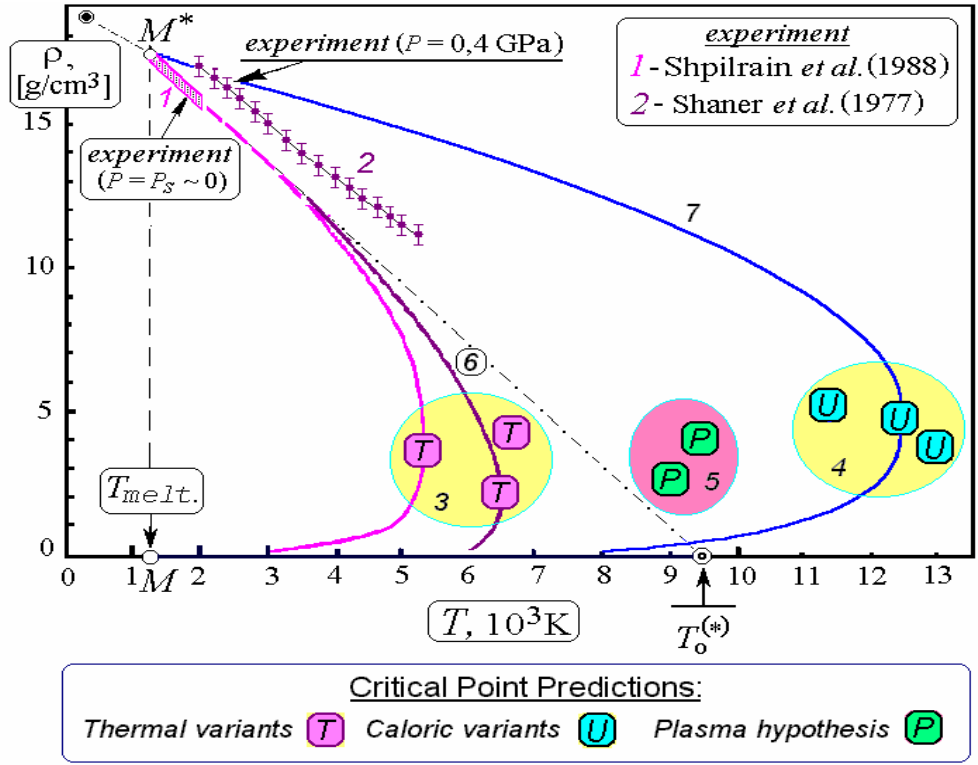

Fig.1. Phase diagram of Uranium *(density-temperature)*

*Experimental data*: *1* – [22], *2* – [21, 16];
*Critical point predictions*: *3* – thermal variants [15-17], *4* – caloric variants [10–14], *5* – «plasma» hypothesis [8,9]
*Phase boundary predictions*: *6* – cesium-like reconstruction [17,19]; *7* – calculations via "chemical model" consolidated with caloric EOS calibration {code SAHA-IV [6,7]}

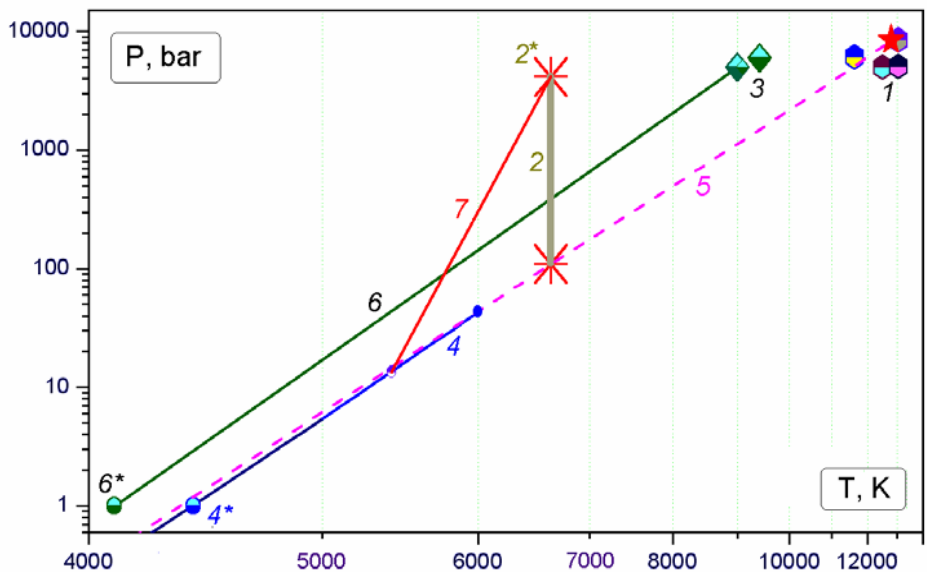

Fig.2. Critical Point and Vapor Pressure of Uranium

*Critical point predictions*: *1* – based on vaporization heat (caloric EOS) [10-14], *2* – based on experimentally measured thermal expansion of liquid (thermal EOS) [15–17]: *2\** – [15,16]; *3* – based on "plasma" hypothesis [8,9];

*Saturation curve*: *4* – IVTAN-Handbook recommended ($T < 6000$) [24,25] (*4\** – boiling point); *5* – $P_S$ ($T_S$) calculated via chemical model consolidated with caloric EOS calibration {code SAHA-IV [6,7]} (asterisk – critical point); *6* – H. Hess recommended hypothetical saturation curve $P_S$ ($T_S$) [9] compatible with "plasma" approach [8] (*6\** – boiling point); *7* – hypothetical anomalous high-temperature extrapolation of saturation curve compatible with D.Young's critical point prediction [15] and experimentally supported [23] IVTAN-Handbook recommendations [24,25] ($T < 5400$ K).

The resolution of the problem of true critical point parameters for uranium and some other "bad" metals [19, 20] (as well as for uranium dioxide, $UO_2$ [6, 18]) is open question at the moment. The problem may be effectively solved by arranging new *benchmark experiment*. The well-controlled quasi-uniform volume heating under the Heavy Ion Beams (HIB) irradiation seems to be especially promising for this purpose. It should be stressed [1] that one should change essentially his priorities in HIB development when he plan to use HIB irradiation for study of thermophysical properties. Development of new regimes of HIB heating is needed for this purpose. In particular, regime of *quasi-isobaric* heating as well as the regime of *saturation curve tracing* are desirable especially in the case of critical point problem investigation with HIB. In frames of such application of HIB the latter could became an uncompetitive tool for study of phase transition phenomenon for a wide number of materials with high-temperature location of critical point.

### HIB in Resolution of Uranium Critical Point Problem

As far as the uranium critical point problem is concerned it was stressed [17] that in search of resolution of the problem one is impelled to *disavow* the results of one (or more) existing experiments or/and to assume at least one (or more) anomaly in properties of high-temperature part of uranium gas-liquid phase transition. Similar situation is valid for problem of CP of non-congruent vaporization in uranium dioxide [18]. In both the cases new experiments with application of HIB can be deciding [1].

In accordance with [17, 20] there may be following variants of possible explanation for mentioned above strong contradiction in different predictions of uranium CPP:

**A** – *Wrong experimental base*
- (A1) – Rough and simultaneous mistake in both measurements of density of isobarically expanded liquid uranium [21, 22].
- (A2) – Rough mistake in liquid uranium $Cp$ measurements up to $T \leq 5400$ K [23] which were put into the base for IVTAN-Handbook recommendations [24, 25] for the uranium saturation pressure data, $P_s(T_s)$.
- (A3) – Rough mistake in the experimentally known data on evaporation heat, $\Delta H_s(T_s)$, which are recommended in IVTAN-Handbook [25]

To check the (A1) version one should realize HIB application in the regimes of *quasi-isobaric* heating *or/and* in the regime of *boiling curve tracing,* both the regimes being consolidated with measurement of liquid density change under heating, $\rho_{liquid}(T)$, in temperature interval $T \sim T_m \div 0.5\text{-}0.7\, T_c = 1.4 \div 7\, 10^3 \text{K}$.

To check the (A2) version one should realize the same non-traditional regimes of HIB application but consolidated with measurement of liquid uranium heat capacity, $C_P$, *or/and* with direct measurement of vapor pressure $P_s(T_s)$.

**B** – *Anomalies in the high-temperature part of uranium gas-liquid phase boundary.*
If one believes in correctness of all mentioned above experimental data he is forced to seek the resolution of the discussed incompatibility at least in one of following *anomalies*. Namely, one had to find the true version among following ones:
- (B1) – Anomalous (*non-convex*) form of density–temperature coexistence curve at high temperatures ($T > 5'000$ K) – ordinarily *totally convex* for all the materials. The only known exclusion from this rule is *theoretically predicted* [6, 18] non-convex form of $\rho_{liquid}(T)$ in the case of *non-congruent* evaporation in $UO_{2 \pm x}$.
- (B2) – Existing of sharp and anomalous <u>upward non-linearity</u> of saturation curve, $P_s(T_s)$, in ($LogP-1/T$) coordinates at $T \sim 5'000$ K $\div$ 13'000 K (see Fig.2) – ordinarily almost linear up to the close vicinity of CP.
- (B3) – Anomalously low value of critical compressibility factor of Uranium, $(PV/RT)_c \ll 1$ – ordinarily confined in interval $0.1 \div 0.4$
- (B4) – Anomalously high value of the ratio of normal density to the critical one, $(\rho_o/\rho_c) \gg 1$ – ordinarily confined in interval $0.1 \div 0.4$

Again, to check the (B1-B2) versions one should realize the same as in the case (A1-A2) non-traditional HIB application in the regimes of *quasi-isobaric* heating *or/and* in the regime of *boiling curve tracing,* being consolidated with the same measurements in temperature interval $T \sim T_m \div 0.5\text{-}0.7\, T_c = 1.4 \div 7 \cdot 10^3 \text{K}$.

It should be stressed [1] that in both cases (A1-A2) as well as in (B1-B2) one *need not* to evaporate the irradiated condensed sample *totally*, but only its small fraction, in fact. Therefore one needs the HIB energy deposition only for heating of the sample in its condensed state. Thermodynamic calculations in SAHA-IV code [7, 6] as well as the IVTAN-Handbook [25] give rise to following minimal energy deposition which are necessary for heating of condensed uranium (uranium dioxide) up to the level $T \sim 7000$ K [26]

$\Delta H_{7000} \equiv H(T = 7000 \text{ K}) - H(T = 300 \text{ K}) \approx 1.3 \text{ kJ/g}$     (condensed Uranium)

$\Delta H_{7000} \equiv H(T = 7000 \text{ K}) - H(T = 300 \text{ K}) \approx 2.7 \text{ kJ/g}$     (condensed Uranium Dioxide)

Thus, one can conclude that only a scanty HIB energy deposition is needed for resolution of many important applied problems when HIB is being applied to thermophysical investigations

**Heavy Ion Beam Perspectives for Study of Thermophysical Properties**

One should change essentially his priorities in HIB development when he wants to use the HIB for study of thermophysical properties of materials [1].

MAJOR PRIORITIES
- *Uniformity of heating,*
- *Careful control of HIB energy deposition,*

MAJOR PREFERENCE

When one uses the HIB heating in study of thermophysical properties he should strive for its *direct measurement* rather than to withdraw the required thermophysical information from results of joint numerical simulation for combination of thermophysics, hydrodynamics and heat transfer.

OTHER PREFERENCES
- One should work for *relatively low temperatures* of irradiated material rather than for the highest ones: $- T \sim 0{,}3 \div 2$ eV
    (It is equivalent to $\Delta U \sim 2 \div 10$ kJ/g when heating the heavy materials)
- One should work for *relatively low densities* of material rather than for the highest ones:
    $\rho \sim 0{,}1 \div 1{,}0\, \rho_0$
- The *planar geometry* of irradiated sample is preferable rather than other ones.
- One should work for *defocused beam* rather than for that in-focus.
- One should work for *divergent beam* to compensate as much as possible the increasing of HIB stopping power along the beam trajectory. When doing so one could put the sample *behind* the beam crossover rather than to put it just in the beam focus.
- One should *cut-off the Bragg's peak* rather than to use it for energy deposition.

**Irradiation of Highly Dispersed Porous Material – Perspective Way of HIB Application**

Quasi-isobaric regime of HIB heating seems to be realistic when the HIB energy deposition being combined with the use of highly dispersed porous material as irradiating sample. Even more, the regime close to boiling curve tracing in pressure-enthalpy coordinates is perspective when evaporating porous material being heated by proper HIB irradiation.

The main idea [Ios99] is that each single grain within the porous sample to be small enough so that its hydrodynamic time, $\tau_d \equiv d/a_{sound}$, being much shorter than the beam duration time, $\tau_{HIB}$. In its turn the time $\tau_{HIB}$ should be much shorter than the hydrodynamic time of total sample, $\tau_D \equiv D/a_{sound}$, (here $d$ and $D$ are the single grain and the total sample diameters, and $a_{sound}$ is the sample material sound speed). It's presumed that when the inequality $\tau_d \ll \tau_{HIB} \ll \tau_D$ is fulfilled the single grain expansion is *quasi-free* (especially after the grains melting) and corresponding heating regime is quasi-isobaric or boiling curve tracing until the moment when the initial volume of inter-grain voids is totally exhausted due to expansion of heated material. This expansion degree is well-controlled quantity by known initial porosity of the sample, $m \equiv \rho_0/\rho_{00} > 1$. It's presumed also that after the moment of the sample entirety achievement the heating regime became *isochoric* and corresponding significant *pressure jump* should distinctly mark this moment. Fig. 3 below illustrates schematically the discussed idea of new non-standard regimes of HIB application in thermophysical measurements.

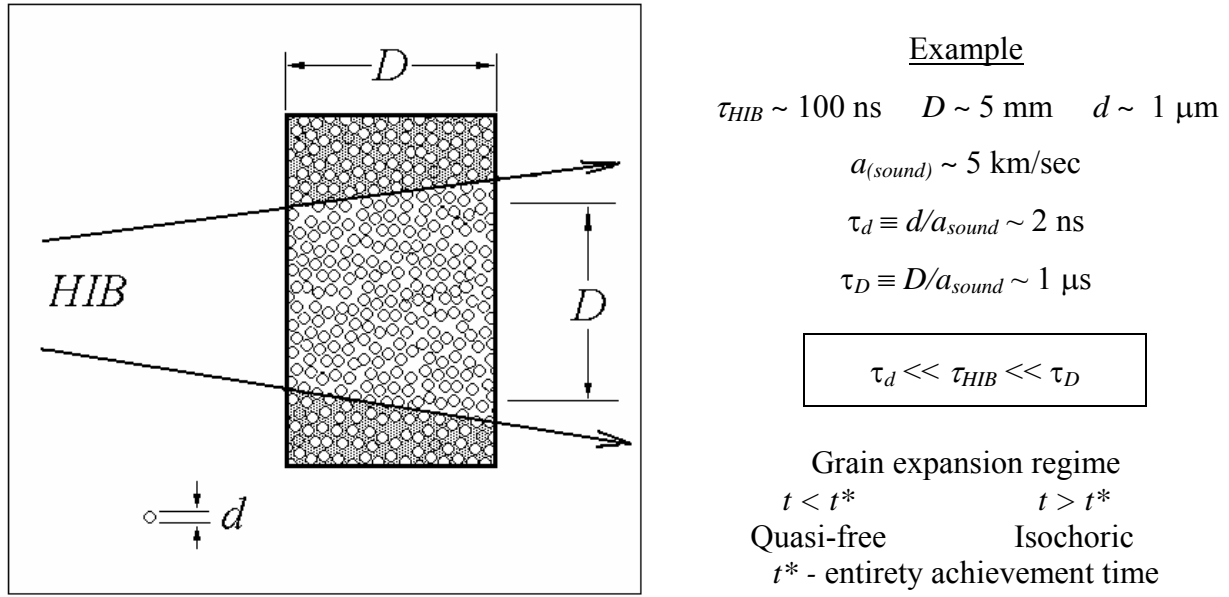

**Figure 3**. HIB irradiation of highly dispersed porous material. Perspective of quasi-isobaric (at $P = const$) or vapor pressure tracing regimes of HIB heating in thermophysical investigations.

Thus, fixation of the moment of such entirety achievement gives the caloric density expansion coefficient of material, $(\partial\rho_{liquid}/\partial H)_P$ or $(\partial\rho_{liquid}/\partial H)_{Boiling}$, when the HIB irradiation being consolidated with careful beam energy deposition control. This output may be enforced when the sample temperature being measured simultaneously or when the heat capacity of irradiated material is known from independent experiment. This is the case for uranium dioxide, $UO_2$. The heat capacity of liquid $UO_2$ was carefully measured up to $T \sim 8000$ K by Ronchi [27]. Therefore the heating by HIB seems to be especially promising as an effective tool for systematic study of so-called *non-congruent* phase transition – striking and mostly unusual sort of high-temperature phase equilibrium in chemically active strongly coupled plasmas. Phase transition in uranium dioxide [6, 18] is remarkable example of this non-congruency.

**Acknowlegements**

The work is supported by Grant CRDF № MO-011-0 and by RAS Scientific Program "Physics and Chemistry of Extremal States of Matter" (2001-2002).